\documentclass[prd,superscriptaddress,amsfonts,amssymb,amsmath,showpacs,twocolumn]{revtex4-2}
\usepackage{bm}
\usepackage{amsfonts}
\usepackage{latexsym}
\usepackage[latin1]{inputenc}
\usepackage{graphicx}
\usepackage{amsmath}
\usepackage{palatino}
\usepackage{mathpazo}
\usepackage{textcomp}
\usepackage{float}
\usepackage{booktabs}
\usepackage{dcolumn}
\usepackage{booktabs}
\usepackage{multirow}
\usepackage{hyperref}
\hypersetup{colorlinks,citecolor=blue}
\usepackage{amsmath}
\usepackage{xcolor}
\usepackage{orcidlink}
\usepackage[caption=false]{subfig}
\usepackage{commath}
\def\jnl@style{\it}
\def\aaref@jnl#1{{\jnl@style#1}}

\def\aaref@jnl#1{{\jnl@style#1}}

\def\aj{\aaref@jnl{AJ}}                   
\def\apj{\aaref@jnl{ApJ}}                 
\def\apjl{\aaref@jnl{ApJ}}                
\def\apjs{\aaref@jnl{ApJS}}               
\def\apss{\aaref@jnl{Ap\&SS}}             
\def\aap{\aaref@jnl{A\&A}}                
\def\aapr{\aaref@jnl{A\&A~Rev.}}          
\def\aaps{\aaref@jnl{A\&AS}}              
\def\mnras{\aaref@jnl{Mon.~Not.~Roy.~Astron.~Soc.}}             
\def\prd{\aaref@jnl{Phys.~Rev.~D}}        
\def\prc{\aaref@jnl{Phys.~Rev.~C}}  
\def\prl{\aaref@jnl{Phys.~Rev.~Lett.}}    
\def\qjras{\aaref@jnl{QJRAS}}             
\def\skytel{\aaref@jnl{S\&T}}             
\def\ssr{\aaref@jnl{Space~Sci.~Rev.}}     
\def\zap{\aaref@jnl{ZAp}}                 
\def\nat{\aaref@jnl{Nature}}              
\def\aplett{\aaref@jnl{Astrophys.~Lett.}} 
\def\apspr{\aaref@jnl{Astrophys.~Space~Phys.~Res.}} 
\def\physrep{\aaref@jnl{Phys.~Rep.}}      
\def\physscr{\aaref@jnl{Phys.~Scr}}       
\def\commat{\aaref@jnl{Comm.~Math.~Phys.}}              
\def\science{\aaref@jnl{Science}}               
\def\cqg{\aaref@jnl{Classical Quant.~Grav.}}            
\def\jpcs{\aaref@jnl{JPCS}}                                     
\def\ijmpd{\aaref@jnl{Int.~J.~Mod.~Phys.~D}}                    
\def\grg{\aaref@jnl{Gen.~Relat.~Gravit.}}               
\def\rpp{\aaref@jnl{Rep.~Prog.~Phys.}}          
\def\npa{\aaref@jnl{Nucl.~Phys.~A}}        
\def\lrr{\aaref@jnl{Living Rev.~Rel.}}                   
\def\jcap{\aaref@jnl{J.~Cosmology Astropart.~Phys.}}    
\def\rmp{\aaref@jnl{Rev.~Mod.~Phys.}}   
\def\epjc{\aaref@jnl{Eur.~Phys.~J.~C}}

\allowdisplaybreaks[1]

\begin{document}

\color{black}       

\title{Cosmic expansion history analysis with Hubble parametrization in $Q^{n}$ gravity}

\author{M. Koussour\orcidlink{0000-0002-4188-0572}}
\email[Email: ]{pr.mouhssine@gmail.com}
\affiliation{Department of Physics, University of Hassan II Casablanca, Morocco.}

\author{N. Myrzakulov\orcidlink{0000-0001-8691-9939}}
\email[Email: ]{nmyrzakulov@gmail.com}
\affiliation{L. N. Gumilyov Eurasian National University, Astana 010008,
Kazakhstan.}

\date{\today}

\begin{abstract}
We propose a specialized parameterization for the Hubble parameter, inspired by $\Lambda$CDM cosmology, to investigate the cosmic expansion history of the Universe. This parameterization is employed to analyze the Universe's late-time behavior within the context of $Q^n$ gravity, where $Q$ represents non-metricity. By using data from 57 Hubble data points, 1048 supernova (SNe) data points, and 6 baryon acoustic oscillation (BAO) data points, we determine the optimal values for the model parameters. Additionally, we explore three distinct cosmological models based on the parameter $n$, specifically when it takes on the values of $0.55$, $1.5$, and $2.0$. The results of our analysis indicate that our proposed parameterization, along with the associated models for different values of $n$, predicts an accelerated cosmic expansion phase.
\end{abstract}

\maketitle

\section{Introduction}

\label{sec1}

The theory of General Relativity (GR) is widely acknowledged as the prevailing explanation for gravity and has successfully withstood numerous experimental and observational examinations. Notably, the detection of gravitational waves, as per the framework of GR, has been achieved through the LIGO and Virgo detectors, as documented in Refs. \cite{A1, A2, A3, A4, A5}. Despite its numerous appealing aspects and this remarkable achievement, there remain several theoretical obstacles that prompt us to seek certain improvements in GR. Additionally, GR does not provide sufficient concepts to address issues such as initial singularity, fine-tuning, flatness problems, cosmological constant, and cosmic coincidence problems \cite{G1, G2, G3, G4}. In the literature, numerous alternative theories have been put forth to tackle these challenges. These theories have been examined for their relevance in understanding the behavior of the expanding Universe, as evidenced by Refs. \cite{E1, E2, E3, E4, E5, E6, E7, E8, E9, E10, E11}. They primarily involve alterations to the gravitational portion of the Einstein-Hilbert action. Conversely, adjustments to the matter component of the Einstein-Hilbert action give rise to dynamic models, including quintessence, k-essence, Chaplygin gas, and holographic Dark Energy (DE) models, as referenced in \cite{M1, M2, M3, M4, M5}. These updated models have demonstrated their capacity to effectively explain the present-day accelerated expansion of the Universe, substantiated by a plethora of experimental observations \cite{O1, O2, O3, O4, O5, O6, O7, O8, O9, O10}.

Among the foundational alternative theories, one theory that distinguishes itself is $f(R)$ gravity, which permits the construction of cosmological frameworks where curvature invariants vary as functions of the Ricci scalar $R$. Research into $f(R)$ gravity models \cite{R1, R2} has proven to be particularly suitable for investigating the unification of early-time inflation and late-time acceleration. Existing literature underscores the crucial role of higher-order curvature terms within the $f(R)$ gravity model in circumventing cosmological singularities \cite{R3, R4, R5}. Furthermore, Harko et al. \cite{T1} introduced the $f(R,\mathcal{T})$ gravity theory as an extension of $f(R)$ gravity. This extended theory incorporates an arbitrary function involving not only the Ricci scalar $R$ but also the trace of the energy-momentum tensor $\mathcal{T}$ \cite{T2, T3}.

In this paper, we will adopt the recently formulated $f(Q)$ gravity theory \cite{J.B.}, where the symbol $Q$ represents the non-metricity scalar. In the realm of gravitational physics, three distinct geometrical variables-curvature, torsion, and non-metricity-are utilized to elucidate gravitational effects in the space-time manifold. In the framework of GR, gravitational phenomena are primarily associated with space-time curvature. Conversely, torsion and non-metricity offer alternative but equivalent representations of GR and its associated gravity theories. These are referred to as Teleparallel Gravity (TG) and Symmetric Teleparallel Gravity (STG), respectively, and both are on par with GR. The $f(R)$ gravity extends GR by focusing on curvature and disregarding torsion and non-metricity \cite{R1, R2, R3, R4, R5}. Similarly, the $f(T)$ gravity is an extension of TG, emphasizing torsion while neglecting non-metricity and curvature \cite{Q1}. Lastly, the $f(Q)$ gravity theory serves as an extension of STG, with its primary focus on non-metricity and exclusion of torsion and curvature. In the context of STG, non-metricity quantifies variations in the length of vectors during teleparallel transport. This approach not only enables the covariantization of conventional GR coordinate calculations but also offers a valuable geometric representation. In STG, the associated energy-momentum density essentially corresponds to the Einstein pseudo tensor, which, within this geometric framework, transforms into a genuine tensor. Despite being a relatively recent theory, $f(Q)$ gravity has already demonstrated its potential through several intriguing applications in the literature. The first cosmological solutions within the framework of $f(Q)$ gravity are presented in Refs. \cite{Q2, Q3}, while Refs. \cite{Q4, Q5} delve into $f(Q)$ cosmography and the study of energy conditions. Furthermore, an investigation into cosmological perturbations in $f(Q)$ gravity was undertaken in \cite{Q2}. This study revealed that tensor perturbations exhibit a rescaling of the corresponding Newton constant, while vector perturbations do not contribute in the absence of vector sources. In the scalar sector, two additional propagating modes were identified, indicating that $f(Q)$ theories introduce, at a minimum, two extra degrees of freedom. Additionally, a power-law model has been explored within the context of quantum cosmology \cite{Q6}. Furthermore, cosmological solutions and the growth indices of matter perturbations have been analyzed for a polynomial functional form of $f(Q)$ \cite{Q3}. Notably, Harko et al. investigated matter coupling in $f(Q)$ gravity, assuming a power-law function \cite{Q8}. These diverse studies underscore the wide-ranging applicability and potential of the $f(Q)$ gravity theory in various cosmological and gravitational scenarios.

In addition to the cosmological constant, which maintains a constant energy density for DE across the Universe, there exist various methodologies for modeling the evolution of DE \cite{Pacif1, Pacif2}. One highly effective approach involves creating parametric descriptions of cosmological parameters as functions of time or redshift. These parameters encompass critical factors such as the Hubble parameter, deceleration parameter, and equation of state parameter \cite{H1, H2, DP1, DP2, DP3, DP4, DP5, DP6, DP7}. These parametric models are then compared to observational data to evaluate their consistency with the cosmos as we observe it. However, it's important to note that such models exhibit better alignment with current observational constraints only for specific sets of model parameters, and the quest for a robust cosmological explanation for DE persists. In light of these considerations, we have adopted a unique form of the Hubble parameter, closely resembling the well-accepted $\Lambda$CDM model in cosmology, wherein the deceleration parameter gives rise to an accelerated phase for the Universe. Section \ref{sec3} provides a detailed exploration of the characteristics of this parametrization. Furthermore, we have utilized data from the Hubble parameter, Type Ia supernovae (SNe), and Baryon Acoustic Oscillations (BAO) to impose constraints on the model parameters inherent to our parametric representation.

The structure of this document is as follows: In Sec. \ref{sec2}, we provide an introduction to the fundamental concepts of symmetric teleparallel $f\left( Q\right) $ gravity. Sec. \ref{sec3} delves into the dynamics of background equations for a flat FLRW cosmology. The $Q^{n}$ gravity model is examined in detail in Sec. \ref{sec4}, where we derive equations for density, equation of state, and deceleration parameter relevant to our chosen parametrization. In Sec. \ref{sec5}, we utilize data from 57 points sourced from Hubble data, 6 points from baryon acoustic oscillation data, and 1048 points from Type Ia supernovae to place constraints on the model parameters. Furthermore, Sec. \ref{sec6} investigates the physical behavior of cosmological parameters, including the deceleration parameter, energy density, and equation of state parameter, corresponding to the constrained values for the model parameters. Finally, our findings are elucidated in Sec. \ref{sec7}.

\section{$f(Q)$ formalism}

\label{sec2}

This section offers an insight into the STG formalism, its extension to encompass a $f(Q)$ function, and the ensuing cosmological implications. In STG, gravity is characterized by a distinctive approach where the primary geometric variable dictating the properties of gravitational interaction is represented by the non-metricity scalar $Q$. This scalar geometrically encapsulates the variations in the length of a vector during parallel transport, as opposed to the curvature $R$ as employed in GR. Consequently, gravity is characterized by a non-zero non-metricity, while both the Riemann and torsion tensors register as zero within the ST representation. This approach leads to the substitution of the Levi-Civita connection with the ST connection during the derivation of the field equations \cite{J.B.}. The connection between the metric tensor $g_{\mu \nu }$ of GR and non-metricity is articulated as \cite{Xu},
\begin{equation}
Q_{\gamma \mu \nu }=-\nabla _{\gamma }g_{\mu \nu }=-\frac{\partial g_{\mu
\nu }}{\partial x^{\gamma }}+g_{\nu \sigma }\overline{\Gamma }_{~\mu \gamma
}^{\sigma }+g_{\sigma \mu }\overline{\Gamma }_{~\nu \gamma }^{\sigma }.
\label{2f}
\end{equation}

In the ST representation, the general affine connection $\overline{\Gamma }%
_{\ \mu \nu }^{\gamma }$ is related to the Riemannian part as \cite{Tom},%
\begin{equation}
\overline{\Gamma }_{\ \mu \nu }^{\gamma }=\Gamma _{\ \mu \nu }^{\gamma
}+L_{\ \mu \nu }^{\gamma },
\end{equation}%
with%
\begin{equation}
\Gamma _{~\mu \nu }^{\gamma }=\frac{1}{2}g^{\gamma \sigma }\left( \frac{%
\partial g_{\sigma \nu }}{\partial x^{\mu }}+\frac{\partial g_{\sigma \mu }}{%
\partial x^{\nu }}-\frac{\partial g_{\mu \nu }}{\partial x^{\sigma }}\right)
,
\end{equation}%
and 
\begin{equation}
L_{~\mu \nu }^{\gamma }=\frac{1}{2}g^{\gamma \sigma }\left( Q_{\nu \mu
\sigma }+Q_{\mu \nu \sigma }-Q_{\gamma \mu \nu }\right) =L_{~\nu \mu
}^{\gamma },
\end{equation}%
represent the Levi-Civita connection of the metric tensor $g_{\mu \nu }$ and
the disformation tensor, respectively. Contraction of the non-metricity
tensor leads to%
\begin{equation}
Q\equiv -g^{\mu \nu }\left( L_{\ \ \beta \mu }^{\alpha }L_{\ \ \nu \alpha
}^{\beta }-L_{\ \ \beta \alpha }^{\alpha }L_{\ \ \mu \nu }^{\beta }\right) .
\end{equation}

The ST gravity is a geometric explanation of gravity that is fully
equivalent to GR. This correspondence (or equivalence) is easily
demonstrated in the so-known coincident gauge, about which $\overline{\Gamma 
}_{\ \mu \nu }^{\gamma }=0$. Therefore, by applying the constraint that the
connection be symmetric, the Levi-Civita connection may be represented in
terms of the disformation tensor as $\Gamma _{\ \mu \nu }^{\gamma }=-L_{\
\mu \nu }^{\gamma }$. There are certain significant differences between the
ST and GR models. Because the curvature tensor vanishes under ST gravity,
the total geometry of space-time is flat. As a result, the global geometry
is also of the Weitzenback type \cite{Xu}. Furthermore, gravitational forces
occur due to variations in the length of the vector itself (non-metricity),
rather than the rotation of the angle between two vectors in parallel
transport. Further, similar to $f(R)$ gravity, the action may be generalized
by adding an arbitrary function of the non-metricity scalar $f(Q)$ to the
gravitational Lagrangian, such that the action yields 
\begin{equation}
S=\int {d^{4}x\tau }\left( \frac{1}{2}{f(Q)+L_{m}}\right) .  \label{2k}
\end{equation}

Here, $L_{m}$ denotes the Lagrangian density of matter and $\tau $ denotes $%
\tau =det(g_{\mu \nu })=\sqrt{-g}$. Similarly to $f(R)$ gravity, ${f(Q)}$
will be accountable for the deviation from GR, where, for example, if the
function is assumed to be constant, we may duplicate the $\Lambda $CDM
model. We can only extract two independent traces from the non-metricity
tensor $Q_{\gamma \mu \nu }$, because of the symmetry of the metric tensor $%
g_{\mu \nu }$,%
\begin{equation}
Q_{\gamma }\equiv Q_{\gamma \ \ \ \mu }^{\ \ \mu },\ \ \ \ \tilde{Q}_{\gamma
}\equiv Q_{\ \ \gamma \mu }^{\mu }.  \label{2l}
\end{equation}

The non-metricity conjugate tensor is also defined by 
\begin{eqnarray}
\hspace{-0.5cm} &&P_{\ \ \mu \nu }^{\gamma }\equiv \frac{1}{4}\bigg[-Q_{\ \
\mu \nu }^{\gamma }+2Q_{\left( \mu \ \ \ \nu \right) }^{\ \ \ \gamma
}+Q^{\gamma }g_{\mu \nu }-\widetilde{Q}^{\gamma }g_{\mu \nu }  \notag \\
\hspace{-0.5cm} &&-\delta _{\ \ (\mu }^{\gamma }Q_{\nu )}\bigg]=-\frac{1}{2}%
L_{\ \ \mu \nu }^{\gamma }+\frac{1}{4}\left( Q^{\gamma }-\widetilde{Q}%
^{\gamma }\right) g_{\mu \nu }-\frac{1}{4}\delta _{\ \ (\mu }^{\gamma
}Q_{\nu )}.\quad\quad
\end{eqnarray}

The scalar of non-metricity is obtained by \cite{LZ} 
\begin{eqnarray}
&&Q=-Q_{\gamma \mu \nu }P^{\gamma \mu \nu }=-\frac{1}{4}\big(-Q^{\gamma \nu
\rho }Q_{\gamma \nu \rho }+2Q^{\gamma \nu \rho }Q_{\rho \gamma \nu }  \notag
\\
&&-2Q^{\rho }\tilde{Q}_{\rho }+Q^{\rho }Q_{\rho }\big).
\end{eqnarray}

Furthermore, for the cosmological matter content, the stress-energy momentum
tensor is defined by%
\begin{equation}
\mathcal{T}_{\mu \nu }\equiv -\frac{2}{{\tau }}\frac{\delta ({\tau }L_{m})}{%
\delta g^{\mu \nu }}.  \label{2o}
\end{equation}

The action in Eq. \eqref{2k} is then varied with regard to the metric
tensor components. As a result, we get%
\begin{eqnarray}
\delta S &=&\int {\frac{1}{2}}\delta \left[ {f(Q){\tau }}\right] {{d^{4}x}%
+\int }\delta \left[ {L_{m}{\tau }}\right] {{d^{4}x,}} \\
&=&\int {\frac{1}{2}}\left[ {-}\frac{1}{2}fg_{\mu \nu }\tau \delta g^{\mu
\nu }+f_{Q}\tau \delta Q{-}\mathcal{T}_{\mu \nu }\tau \delta g^{\mu \nu }%
\right] {{d^{4}x,}}  \notag
\end{eqnarray}%
where%
\begin{equation}
\delta Q=2P_{\gamma \nu \rho }\nabla ^{\gamma }\delta g^{\nu \rho }-\left(
P_{\mu \gamma \beta }Q_{\nu }^{\ \ \gamma \beta }-2Q_{\ \ \ \ \mu }^{\gamma
\beta }P_{\gamma \beta \nu }\right) \delta g^{\mu \nu },
\end{equation}

Thus, we derive the field equations of the $f(Q)$ gravity theory by equating
the variation of the gravitational action to zero: 
\begin{widetext}
\begin{equation}\label{2p}
\frac{2}{\sqrt{-g}}\nabla_\gamma (\sqrt{-g}f_Q P^\gamma\:_{\mu\nu}) + \frac{1}{2}g_{\mu\nu}f+f_Q(P_{\mu\gamma\beta}Q_\nu\:^{\gamma\beta} - 2Q_{\gamma\beta\mu}P^{\gamma\beta}\:_\nu) = -T_{\mu\nu}.
\end{equation}
\end{widetext} 

For the purpose of simplicity, we designate $f_{Q}=\frac{df}{dQ}$.

Moreover, after varying the gravitational action with regard to the
connection, we get the field equations, 
\begin{equation}
\nabla _{\mu }\nabla _{\nu }(\sqrt{-g}f_{Q}P^{\mu \nu }{}_{\gamma })=0
\label{2q}
\end{equation}

It is crucial to emphasize that Eq. (\ref{2q}) remains applicable solely under the condition of the absence of hypermomentum. Furthermore, the Bianchi identity establishes that this equation is automatically fulfilled when the metric equations of motion are satisfied, as cited in \cite{Jim4}.

\section{Background dynamics in $f(Q)$ gravity}

\label{sec3}

To investigate the cosmological consequences of $f(Q)$ gravity in the setting of a homogeneous, isotropic, and spatially flat Universe, we observe
that this corresponds to the FLRW geometry denoted by the line element \cite%
{Ryden},%
\begin{equation}
ds^{2}=-dt^{2}+a^{2}(t)\delta _{ij}dx^{i}dx^{j},  \label{3a}
\end{equation}%
where the scale factor $a(t)$ is a measure of cosmological expansion at a cosmic time $t$. The stress-energy momentum tensor that characterizes
the matter content of the Universe is%
\begin{equation}
\mathcal{T}_{\mu \nu }=(\rho +p)u_{\mu }u_{\nu }+pg_{\mu \nu }  \label{3c}
\end{equation}%
where $\rho $\ and $p$ representing the energy density and pressure derived
from the global stress-energy momentum tensor, respectively. In addition, $%
u^{\mu }=(1,0,0,0)$ are components of the four velocities of the ideal
cosmological fluid. So, from \eqref{2p}, the modified Friedmann-like
equations describe the dynamics of the Universe for the function $f(Q)$\
become \cite{LZ, Q8},%
\begin{equation}
3H^{2}=\frac{1}{2f_{Q}}\left( -\rho +\frac{f}{2}\right) ,  \label{3d}
\end{equation}%
and%
\begin{equation}
\dot{H}+3H^{2}+\frac{\dot{f}_{Q}}{f_{Q}}H=\frac{1}{2f_{Q}}\left( p+\frac{f}{2%
}\right) ,  \label{3e}
\end{equation}%
where the dots $(\overset{.}{})$ denote derivatives with respect to cosmic
time $t$ and $H\equiv \frac{\overset{.}{a}}{a}$ is the Hubble parameter. The
first equation, known as the Friedmann equation, relates the Universe's
expansion rate to its energy density, whereas the second equation, known as
the acceleration equation, explains how the Universe's rate of expansion
evolves over time in $f(Q)$ gravity. Also, we can see that if $f(Q)=-Q$, the
ST description is similar to GR, however, the dynamics may be completely
changed by assuming a different $f(Q)$ function, because $Q=6H^{2}$ in the
FLRW geometry \cite{LZ}. For this case, Eqs. (\ref{3d}) and (\ref{3e}) reduce to the
Friedmann equations of GR, $3H^{2}=\rho $, and $2\dot{H}+3H^{2}=-p$,
respectively.

The expressions for energy density $\rho $ and isotropic pressure $p$ are
obtained from Eqs. \eqref{3d} and \eqref{3e}, respectively%
\begin{equation}
\rho =\frac{f}{2}-6H^{2}f_{Q},  \label{3i}
\end{equation}%
and%
\begin{equation}
p=\left( \overset{.}{H}+\frac{\overset{.}{f_{Q}}}{f_{Q}}H\right) \left(
2f_{Q}\right) -(\frac{f}{2}-6H^{2}f_{Q}).  \label{3j}
\end{equation}

Thus, the equation of state (EoS) parameter is
\begin{equation}
\omega =\frac{p}{\rho }=-1+\frac{\left( \overset{.}{H}+\frac{\overset{.}{%
f_{Q}}}{f_{Q}}H\right) \left( 2f_{Q}\right) }{\left( \frac{f}{2}%
-6H^{2}f_{Q}\right) }.  \label{3l}
\end{equation}

The modified Friedmann equations, as previously elucidated, comprise merely two distinct equations, each involving four variables: $\rho$, $p$, $H$, and $f$. To comprehensively resolve this system and scrutinize the evolution of the energy density, isotropic pressure, and the EoS parameter over time, an additional pair of constraint equations is imperative. In the subsequent section, we will tackle these equations by applying assumptions drawn from the existing literature.

\section{Cosmological $Q^{n}$ Model}
\label{sec4}

In this paper, we propose a specific power-law model for the $f(Q)$ function, which is expressed as
\begin{equation}
f(Q)=-Q^{n},  \label{4a}
\end{equation}%
where $n$ is a real number that plays a crucial role in shaping the behavior of the cosmological model described by the $f(Q)$ function. Various values of $n$ can give rise to distinct cosmological scenarios, including varying rates of cosmic expansion, early universe inflationary behavior \cite{Inf}, and the nature of late-time accelerated expansion \cite{KOU}. The motivation for the negative sign in the function $f(Q)$ can be attributed to two significant factors. Firstly, we aimed to ensure that our proposed model encompasses GR as a special case when $n=1$. GR is a well-established theory of gravitation, and it was crucial for us to recover its predictions within the framework of our model under certain conditions. However, it does not align with the $\Lambda$CDM model, as the cosmological constant is absent in this case. Secondly, the choice of a positive sign in the action (Eq. (\ref{2k})) influenced the subsequent consideration of the negative sign in the function $f(Q)$ \cite{Q2}. This particular combination of signs leads to intriguing and physically meaningful consequences, which align well with the observed phenomena we aimed to explain within our research. In addition, this feature of the model offers a potential solution to the cosmological constant problem, which has been a longstanding challenge in cosmology. The model exhibits intriguing behavior as the redshift increases, showing only small deviations from the $\Lambda$CDM model. This aspect has been extensively investigated \cite{P1} and highlights the model's ability to effectively describe the late-time universe acceleration while remaining compatible with observational constraints \cite{P2,KOU}. Notably, the model's predictions are consistent with the bounds set by Big Bang Nucleosynthesis (BBN) limitations \cite{Anagnostopoulos}, indicating its ability to account for the early universe's nucleosynthesis and preserving the successes of standard cosmology during those crucial stages.

Then, we propose a simple DE parametrization that shows dynamical behavior
with Universe evolution, similar to the $\Lambda $CDM model. Our principal
objective is to test this idea using present cosmological data. In addition,
there are other interesting works where they employed parameterized
techniques to investigate the nature of DE \cite{Pacif1, Pacif2, H1, H2, DP1,
DP2, DP3, DP4, DP5, DP6, DP7}. Please also see some work on the parametrization approach in the context of modified gravity \cite{Par1, Par2, Par3, Par4, Par5}. G. N. Gadbail et al. reconstruct cosmic
evolution in the $f(Q)$ gravity using the Hubble parameter $\Lambda $CDM
cosmology \cite{G1}. Here, we present the functional form for the evolution
of $H\left( z\right) $ in terms of redshift $z$\ given by

\begin{equation}
H\left( z\right) =\frac{H_{0}}{2}\left[ \sqrt{4\alpha (1+z)^{3}+\beta ^{2}}%
+\beta \right] ,  \label{Hz}
\end{equation}%
\newline
where $H_{0}=H\left( z=0\right) $ is the late time Hubble parameter, $\alpha 
$ and $\beta $ are free model parameters. One important advantage of this
choice is that it similar to the most widely accepted $\Lambda $CDM model in
cosmology i.e. $H\left( z\right) =H_{0}\sqrt{\Omega _{m}^{0}\left(
1+z\right) ^{3}+\Omega _{\Lambda }^{0}}$, where $\Omega _{m}^{0}$ and $%
\Omega _{\Lambda }^{0}$ are the matter density parameter and the
cosmological constant ($\Lambda $) density parameter, respectively \cite{G1}%
. For $\beta =0$, this corresponds to an Universe full of matter with $%
\alpha =\Omega _{m}^{0}$, while for $\alpha =0$, it corresponds to an
Universe filled of DE in the form of a cosmological constant with $\beta =%
\sqrt{\Omega _{\Lambda }^{0}}$. As a consequence, the free parameter $\beta $
is an excellent indicator of deviation of the current DE parametrization
from the $\Lambda $CDM model. Observational data may be used to estimate the
constants $\alpha $ and $\beta $, when $z=0$, we may establish an additional
restriction on the parameters as $\alpha +\beta =1$. In addition, if $\beta
=0$, our Hubble parametrization becomes the power-law. Power-law cosmology
provides an attractive solution to several uncommon problems, including as
flatness and the horizon problem. In literature, the power-law is well
justified \cite{G2}. Kumar examined cosmic parameters using a power-law with
Hubble and SNe data sources \cite{G3}. Rani et al. used statefinder analysis
to investigate the power-law cosmology \cite{G4}.

Another important factor that describes the expansion phase of the Universe is
the deceleration parameter. For this choice of $H\left( z\right) $, the
deceleration parameter $q\left( z\right) $ is given as,

\begin{widetext}
\begin{equation}
q\left( z\right) =-1+\frac{\left( 1+z\right) }{H\left( z\right) }\partial
_{z}H\left( z\right) =-1+\frac{6\alpha (z+1)^{3}}{\beta \left( \beta +\sqrt{%
\beta ^{2}+4\alpha (z+1)^{3}}\right) +4\alpha (z+1)^{3}}.  \label{qz}
\end{equation}

By using Eqs. \eqref{3i}, \eqref{3j} and \eqref{4a}, we obtain%
\begin{equation}
\rho\left( z\right) =2^{n-1}3^{n}(2n-1)H\left( z\right) ^{2n},
\label{4c}
\end{equation}%
\begin{equation}
p\left( z\right) =-6^{n-1}(2n-1)H\left( z\right) ^{2\left( n-1\right)
}\left[ -2n\left( 1+z\right) H\left( z\right) \partial _{z}H\left( z\right)
+3H\left( z\right) ^{2}\right] .
\end{equation}

Again, by using Eqs. \eqref{3l} and \eqref{4a}, we obtain

\begin{equation}
\omega \left( z\right) =-1+\frac{2n}{3}\frac{\left( 1+z\right) }{H\left(
z\right) }\partial _{z}H\left( z\right) =\frac{4\alpha (n-1)(z+1)^{3}-\beta
\left( \beta +\sqrt{\beta ^{2}+4\alpha (z+1)^{3}}\right) }{\beta \left(
\beta +\sqrt{\beta ^{2}+4\alpha (z+1)^{3}}\right) +4\alpha (z+1)^{3}}.
\label{omega}
\end{equation}
\end{widetext}

The given formula of $\omega \left( z\right) $ shows that for $n=0$, this
behaves precisely like the standard $\Lambda $CDM model i.e. $\omega \left(
z\right) =-1$. Thus, the chosen value of the parameter $n$ plays an
essential role in deciding the nature of the cosmological model. Also, Eq. (%
\ref{4c}) indicates that the value of $n$ must satisfy the condition $n\geq 
\frac{1}{2}$ (because $\rho \left( z\right) \geq 0$). So, we will
investigate three possible values of $n$ i.e. $n=0.55$, $n=1.5$ and $n=2.0$.
In the next section, we will attempt to derive the values of the model
parameters using the most recent cosmological data sources.

\section{Observational Data}

\label{sec5}

As more data becomes available, researchers have explored the reconstruction of DE using a variety of observational sources. The primary focus of many studies has been on data from SNe, CMB, and BAO, all of which are valuable for constraining cosmological models. The dataset of the Hubble parameter provides insights into the intricate dynamics of the Universe's expansion. Direct measurements of the Hubble parameter, denoted as $H(z)$ at various redshifts $z$, have been obtained from the ages of the largest, massive, and slowly evolving galaxies. This approach has led to the creation of a novel standard cosmological probe \cite{Jim2}. In this section, we utilize updated Hubble data sources, Pantheon SNe data sources, and BAO data sources to estimate parameter values for our proposed cosmological model. We employ the Markov Chain Monte Carlo (MCMC) method and Bayesian analysis to explore the parameter space of our cosmological model, employing the emcee Python package \cite{Mackey/2013}. It's worth noting that discussions related to the observational data have been presented similarly by several authors \cite{P1, P2, KOU}.

\subsection{Hubble dataset}

The Hubble observations constitute our initial dataset for analysis. It is well-established that the Hubble parameter can directly provide insights into the rate of cosmic expansion. Generally, two widely-used approaches for determining the Hubble parameter at specific redshifts are the Differential Age and the Line of Sight BAO methods. In this investigation, we constrain our model using a set of 57 data points of the Hubble parameter in the redshift range $0.07 \leq z \leq 2.41$, as compiled by Sharov and Vasiliev \cite{Hubble}. For the Hubble data sources, the $\chi^2$ function is represented as,
\begin{equation}
\chi _{Hubble}^{2}=\sum\limits_{k=1}^{57}\frac{[H_{th}(z_{k},H_{0},\alpha
)-H_{obs}(z_{k})]^{2}}{\sigma _{H(z_{k})}^{2}}.
\end{equation}

In this context, we denote $H_{obs}$ as the measured value of the Hubble parameter obtained from cosmic observations, $H_{th}$ as its theoretical estimate, and $\sigma_{H(z_{k})}$ as the corresponding error.

\begin{figure}[]
\includegraphics[width=8.5 cm]{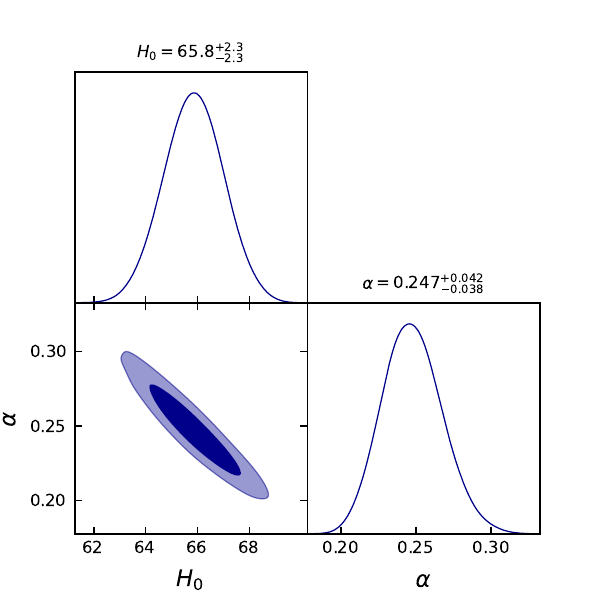}
\caption{Determination of model parameter constraints within the $1-\sigma$ and $2-\sigma$ confidence intervals based on the data from the Hubble data sources.}
\label{C1}
\end{figure}

\begin{figure}[]
\includegraphics[width=8.5 cm]{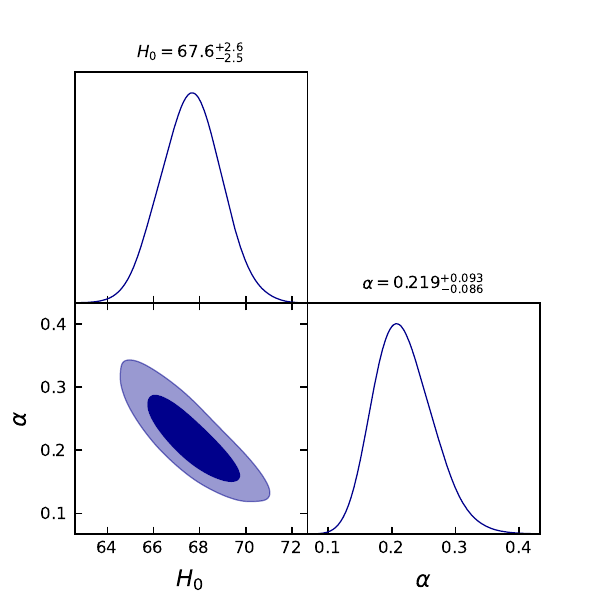}
\caption{Determination of model parameter constraints within the $1-\sigma$ and $2-\sigma$ confidence intervals based on the data from Type Ia SNe sources.}
\label{C2}
\end{figure}

\begin{figure}[]
\includegraphics[width=8.5 cm]{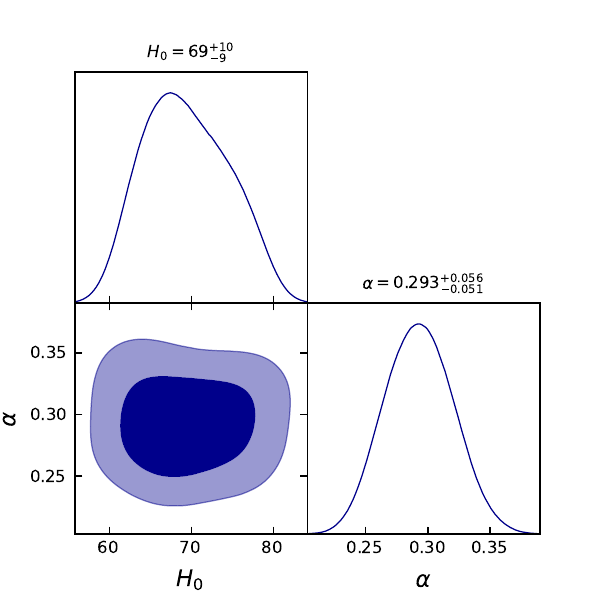}
\caption{Determination of model parameter constraints within the $1-\sigma$ and $2-\sigma$ confidence intervals based on the data from BAO sources.}
\label{C3}
\end{figure}

\subsection{SNe dataset}

In addition, we incorporate the presently accessible Pantheon Type Ia SNe data into our computations. This dataset comprises 1048 SNe samples with observed distance modulus $\mu^{obs}$ within the redshift range of $z\in[0.01,2.3]$ \cite{Scolnic/2018}. For the Pantheon data sources, the $\chi^2$ function is expressed as,
\begin{equation}
\chi _{SNe}^{2}=\sum_{i,j=1}^{1048}\Delta \mu _{i}\left(
C_{SNe}^{-1}\right) _{ij}\Delta \mu _{j},  \label{4b}
\end{equation}%
where, $C_{SNe}$ signifies the covariance matrix \cite{Scolnic/2018}, and 
\begin{equation*}
\quad \Delta \mu _{i}=\mu ^{th}(z_{i},H_{0},\alpha )-\mu
_{i}^{obs},
\end{equation*}%
is the difference between the observed distance modulus value received from
astronomical data and its theoretical values obtained from the model with
the specified parameter $H_{0},\alpha $. Also, the distance modulus is
calculated as $\mu =m_{B}-M_{B}$, where $m_{B}$ and $M_{B}$ denote the
observed apparent magnitude and absolute magnitude at a certain redshift
(Trying to retrieve the nuisance parameter using the new BEAMS with Bias
Correction technique (BBC) \cite{BMS}). Its theoretical value is also
provided by 
\begin{equation}
\mu (z)=5log_{10}\left[ \frac{D_{L}(z)}{1Mpc}\right] +25,  \label{4d}
\end{equation}%
where 
\begin{equation}
D_{L}(z)=c(1+z)\int_{0}^{z}\frac{dy}{H(y,H_{0},\alpha )}.
\end{equation}

\subsection{BAO dataset}

The last constraints in this analysis were given via BAO observation. BAO
investigates oscillations caused in the early Universe by cosmic
perturbations in a fluid comprised of photons, baryons, and dark matter and
linked via Thompson scattering. The Sloan Digital Sky Survey (SDSS), the Six
Degree Field Galaxy Survey (6dFGS), and the Baryon Oscillation Spectroscopy
Survey (BOSS) are some of the BAO observations \cite{BAO1,BAO2}. The
equations used in BAO measurements are, 
\begin{equation}
d_{A}(z)=\int_{0}^{z}\frac{dz^{\prime }}{H(z^{\prime })},  \label{4f}
\end{equation}%
\begin{equation}
D_{V}(z)=\left( d_{A}(z)^{2}z/H(z)\right) ^{1/3},  \label{4g}
\end{equation}%
and 
\begin{equation}
\chi _{BAO}^{2}=X^{T}C_{BAO}^{-1}X,  \label{4h}
\end{equation}%
where $C_{BAO}$ symbolizes the covariance matrix \cite{BAO6}, $d_{A}(z)$ the
angular diameter distance, and $D_{V}(z)$ the dilation scale.

The model parameter constraints are calculated by minimizing the
corresponding $\chi ^{2}$ with MCMC and the emcee library (the results are
shown in Tab. \ref{tab}). Figs. \ref{C1}-\ref{C3} exhibit the $1-\sigma $
and $2-\sigma $ likelihood graphs for the model parameters $H_{0}$ and $%
\alpha $ using Hubble, SNe, and BAO data sources, respectively. The
likelihoods are extremely closely adapted to Gaussian distributions.
Furthermore, Figs. \ref{ErrorHubble} and \ref{ErrorSNe} show the error bar
fitting for both the considered model and the $\Lambda$CDM model. For this analysis, we adopt the specific parameter values obtained from Planck observations, namely, $\Omega _{0}^{m}=0.315$ and $H_0=67.4$ $km/s/Mpc$ \cite{Planck2020}.

\subsection{Bayesian model comparison and selection}

In order to assess the efficacy of our MCMC study, we must perform a statistical evaluation employing the Akaike Information Criterion (AIC) and Bayesian Information Criterion (BIC). The AIC is expressed as \cite{Akaike}
\begin{equation}
    \mathrm{AIC} = \chi^2_{\mathrm{min}}+2d,
\end{equation}
where $\chi^2_{\mathrm{min}}$ is the minimum value of the chi-squared for the model and $d$ is the number of parameters in the model. To compare our results with the standard $\Lambda$CDM model, we use the AIC difference between our model and the standard cosmological model,
\begin{equation}
\Delta\mathrm{AIC}=|\mathrm{AIC}_{\Lambda\mathrm{CDM}}-\mathrm{AIC}_{\mathrm{Model}}|.
\end{equation}

Here,  if the difference ($\Delta\mathrm{AIC}$) is less than 2, it suggests strong support for our model. If the difference is between 4 and 7, there is modest support for our model. However, if the difference is greater than 10, there is essentially no support for our model \cite{Liddle}. Moreover, the BIC is determined by the formula:
\begin{equation}
\mathrm{BIC} =\chi^2_{\mathrm{min}}+d\ln N.
\end{equation}

In this context, $N$ represents the number of data points used in the MCMC analysis. Regarding the BIC, if the difference $(\Delta \mathrm{BIC})$ is less than 2, there is no substantial evidence against the chosen model deviating from the standard $\Lambda$CDM model. If the difference is between 2 and 6, there is evidence against the model. Finally, if the difference is greater than 6, there is strong evidence against our model. Thus, we record the $\chi^2_{\mathrm{min}}$/AIC/BIC data for our model in Tab. \ref{tab}. From these values, we observe that $\Delta\rm AIC=1.2$ for $Hubble$, $1.7$ for $SNe$, and $2.4$ for $BAO$. These findings suggest strong support for all combinations of datasets in favor of the model.

\begin{widetext}

\begin{table*}[h]
\begin{center}
\begin{tabular}{l c c c c c c c c}
\hline\hline 
$Data$              & $H_{0}$ $(km/s/Mpc)$ & $\alpha$ & $\chi^{2}_{min}$ & $AIC$ & $BIC$ \\
\hline
$Priors$   & $(60,80)$  & $(0,1)$ & $-$ & $-$ & $-$\\

$Hubble$ & $65.8_{-2.3}^{+2.3}$ & $0.247_{-0.042}^{+0.038}$ & $30.825$ & $34.825$ & $38.911$ \\

$SNe$   & $67.6_{-2.5}^{+2.6}$ & $0.219_{-0.086}^{+0.093}$ & $1041.996$ & $1051.996$ & $1076.769$ \\

$BAO$   & $69_{-9}^{+10}$ & $0.293_{-0.051}^{+0.056}$ & $4.886$ & $8.886$ & $8.469$\\

\hline\hline
\end{tabular}
\caption{The table presented above displays the best-fit parameters of the model obtained from three distinct datasets: $Hubble$, $SNe$, and $BAO$.}
\label{tab}
\end{center}
\end{table*}

\begin{figure}[h]
\centerline{\includegraphics[scale=0.60]{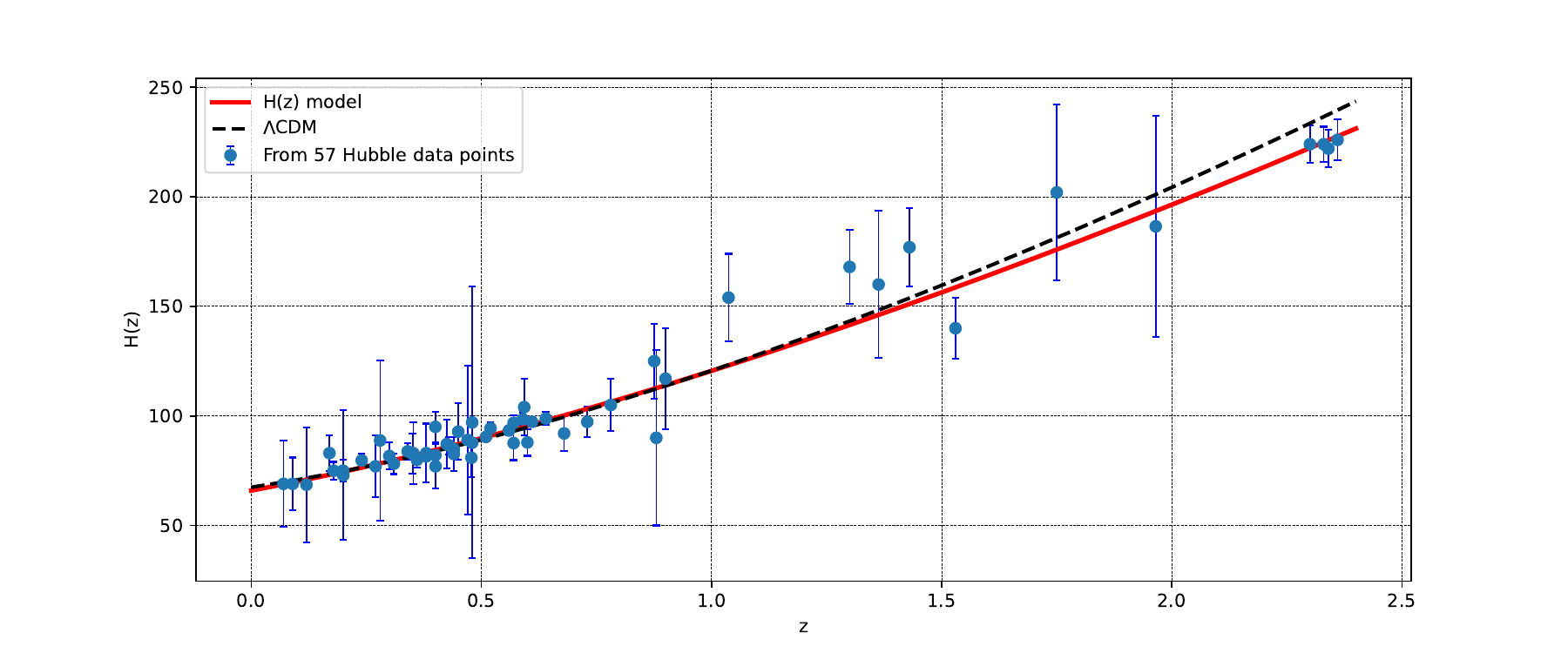}}
\caption{In the plot depicting $H(z)$ against redshift $z$ for our model (depicted in red) and the $\Lambda$CDM model (indicated by black dashed lines), we can observe a well-matched representation for the 57 data points from the Hubble dataset.}
\label{ErrorHubble}
\end{figure}

\begin{figure}[h]
\centerline{\includegraphics[scale=0.60]{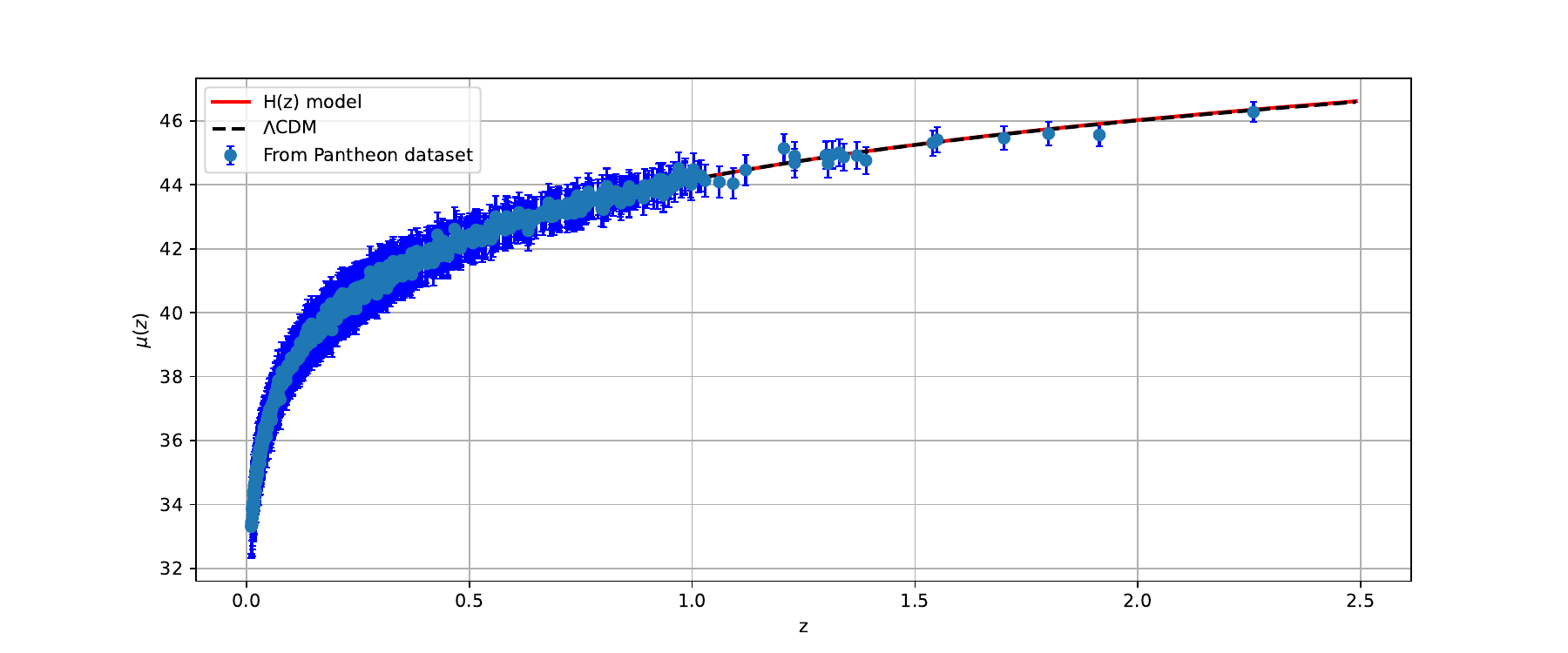}}
\caption{In the plot depicting $\mu(z)$ against redshift $z$ for our model (depicted in red) and the $\Lambda$CDM model (indicated by black dashed lines), we can observe a well-matched representation for the 1048 data points from the Pantheon dataset.}
\label{ErrorSNe}
\end{figure}

\end{widetext}

\section{Consequences of cosmological parameters}

\label{sec6}

The analysis of cosmological parameters has garnered a lot of attention in
modern cosmology to understand the expanding dynamics of the Universe. In this
part, we address cosmological parameters such as deceleration parameter, EoS, speed of
sound, and Om diagnostics for constrained values of model parameters from
Hubble data sources, SNe data sources, and BAO data sources. In addition, we
will investigate the influence of parameter $n$ on the model by examining
three possible values, namely $n=0.55$, $n=1.5$, and $n=2.0$. The evolution
of several cosmological parameters of our $f\left( Q\right) $\ cosmological
model according to the constrained values of the model parameters, is shown
below.

\begin{figure}[tbp]
\includegraphics[scale=0.65]{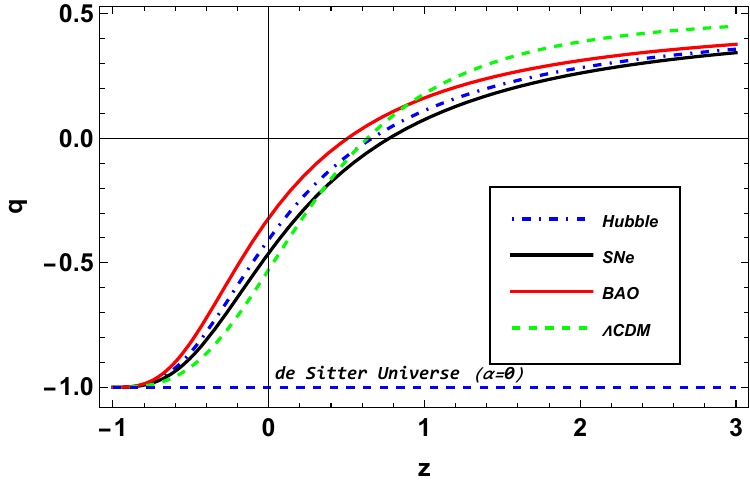}
\caption{Evolution of the deceleration parameter vs redshift $z$. Also, we have included a comprehensive comparison to the well-established $\Lambda$CDM model, we use the latest parameter values derived from Planck observations \cite{Planck2020}.}
\label{Fq}
\end{figure}

\begin{figure}[tbp]
\includegraphics[scale=0.6]{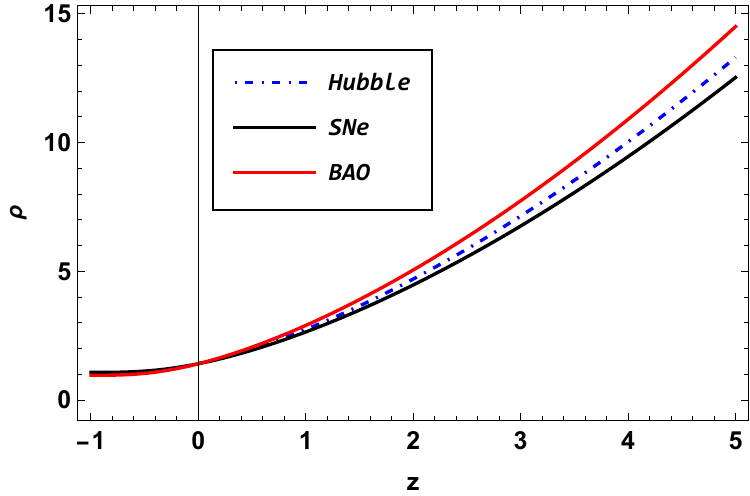}
\caption{Evolution of the energy density of the Universe vs redshift $z$ ($%
n=0.55$).}
\label{Fr0}
\end{figure}

\begin{figure}[tbp]
\includegraphics[scale=0.6]{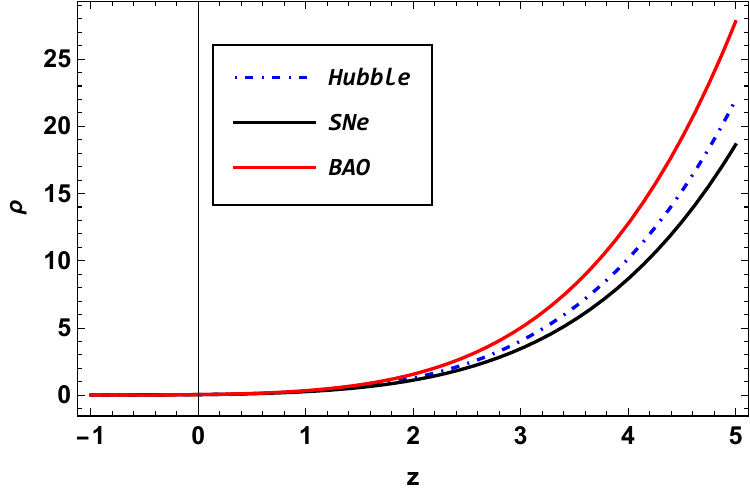}
\caption{Evolution of the energy density of the Universe vs redshift $z$ ($%
n=1.5 $).}
\label{Fr1}
\end{figure}

\begin{figure}[tbp]
\includegraphics[scale=0.6]{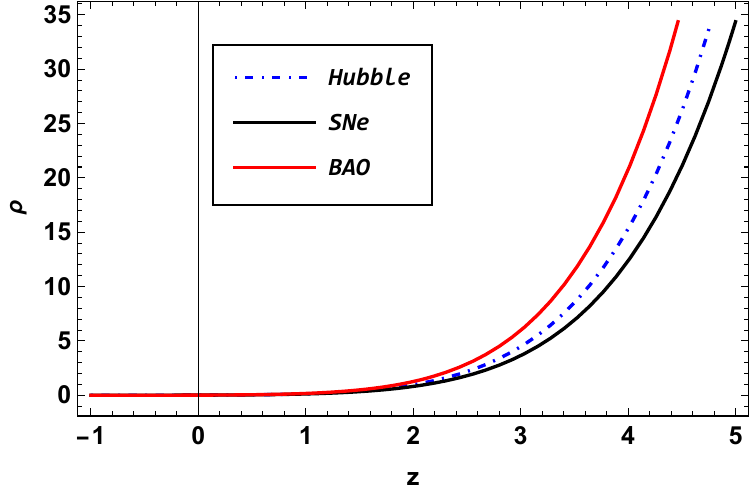}
\caption{Evolution of the energy density of the Universe vs redshift $z$ ($%
n=2.0 $).}
\label{Fr2}
\end{figure}

\begin{figure}[tbp]
\includegraphics[scale=0.6]{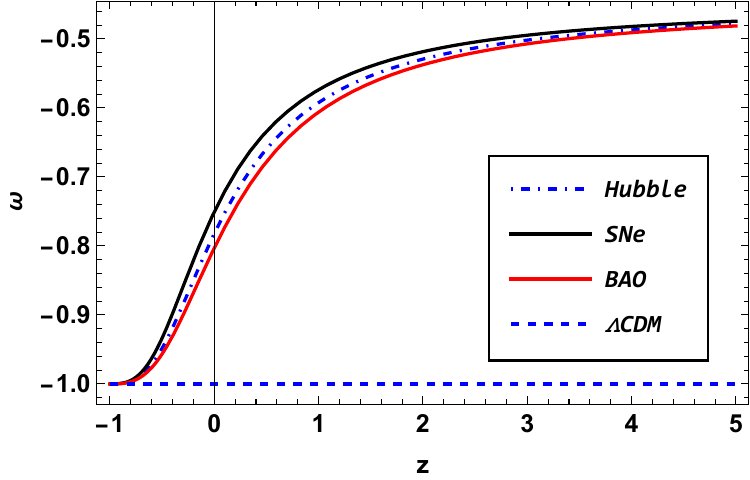}
\caption{Evolution of the EoS parameter of the Universe vs redshift $z$
($n=0.55$).}
\label{FE0}
\end{figure}

\begin{figure}[tbp]
\includegraphics[scale=0.6]{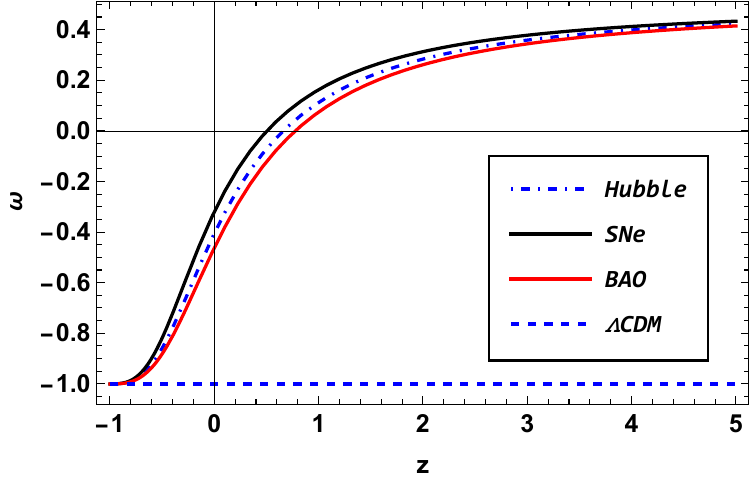}
\caption{Evolution of the EoS parameter of the Universe vs redshift $z$
($n=1.5$).}
\label{FE1}
\end{figure}

\begin{figure}[tbp]
\includegraphics[scale=0.6]{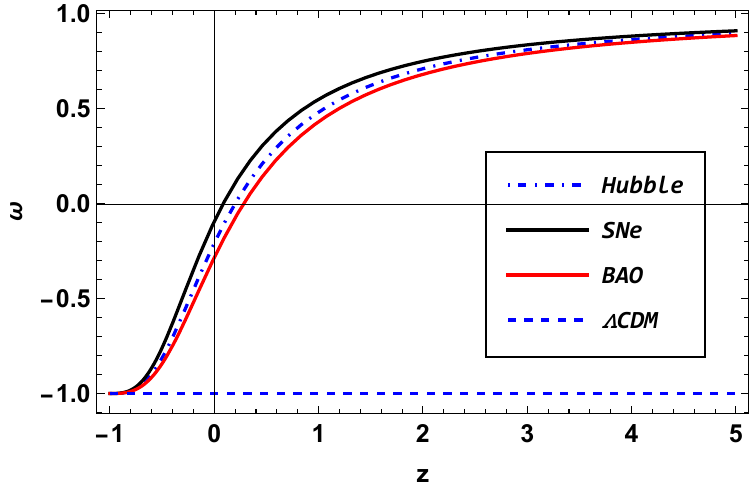}
\caption{Evolution of the EoS parameter of the Universe vs redshift $z$
($n=2.0$).}
\label{FE2}
\end{figure}

\begin{figure}[tbp]
\includegraphics[scale=0.6]{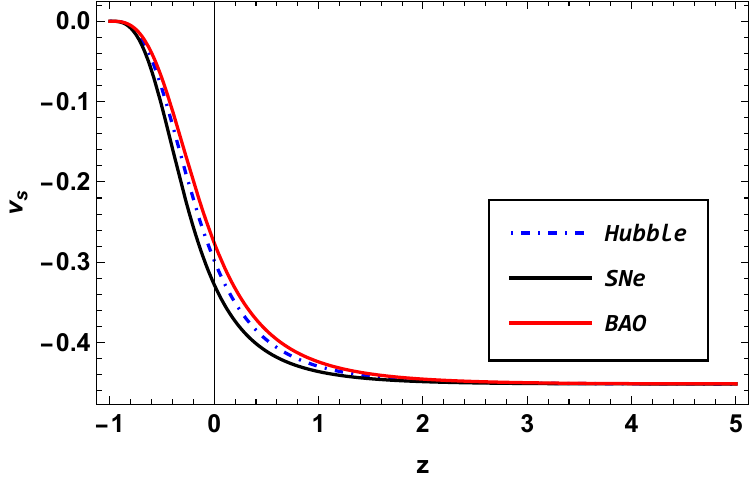}
\caption{Evolution of the speed of sound parameter of the Universe vs
redshift $z$ ($n=0.55$).}
\label{Fv0}
\end{figure}

\begin{figure}[tbp]
\includegraphics[scale=0.6]{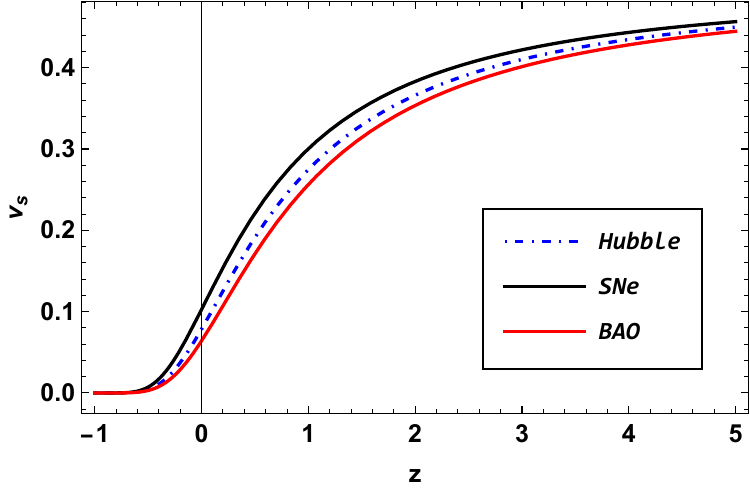}
\caption{Evolution of the speed of sound parameter of the Universe vs
redshift $z$ ($n=1.5$).}
\label{Fv1}
\end{figure}

\begin{figure}[tbp]
\includegraphics[scale=0.6]{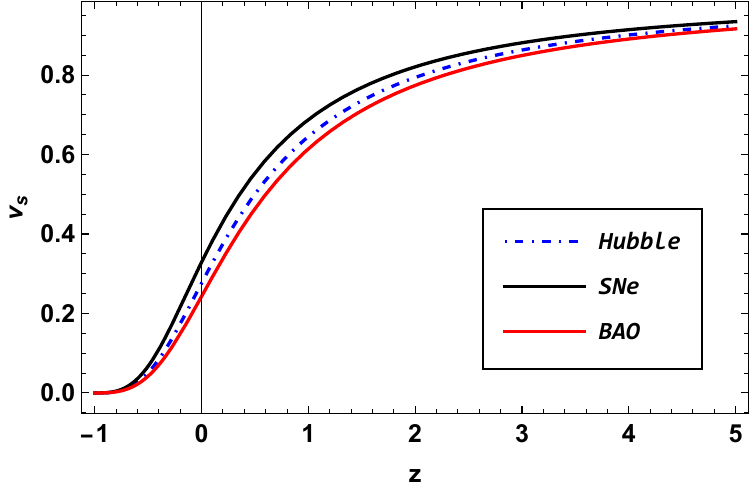}
\caption{Evolution of the speed of sound parameter of the Universe vs
redshift $z$ ($n=2.0$).}
\label{Fv2}
\end{figure}

\begin{figure}[tbp]
\includegraphics[scale=0.6]{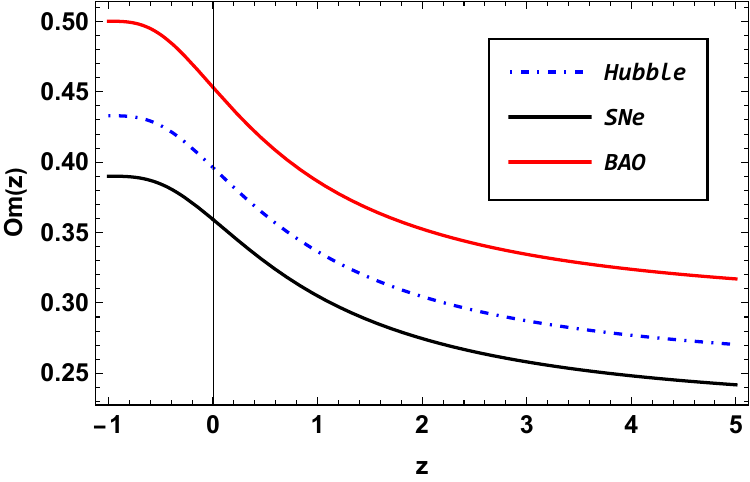}
\caption{Evolution of the Om diagnostic vs redshift $z$.}
\label{FOm}
\end{figure}

The deceleration parameter (DP) is one of the crucial characteristics that must
be considered while discussing the evolution of the Universe. The sign of the
DP (negative or positive) shows if the Universe accelerates or decelerates, it
is expressed as $q=-1-\frac{\overset{.}{H}^{2}}{H^{2}}$. In Eq. (\ref{qz})
we get the DP expression for our model. As can be seen in Fig. \ref{Fq}, the
evolution of DP for our Hubble parametrization exhibits positive behavior
for larger values of $z$, but as the value of $z$ decreases, we obtain
negative behavior of the DP. This suggests that DP for our model represents
a decelerated phase of the Universe in early periods, whereas lowering $z$
results in an accelerated phase of the Universe. The present value of the DP
is $q_{0}=-0.406^{+0.079}_{-0.072}$, $q_{0}=-0.461^{+0.174}_{-0.163}$, and $q_{0}=-0.320^{+0.096}_{-0.088}$ according to the
Hubble, SNe and BAO data sources, respectively. Also, we can see that our
Hubble parametrization leads to a de Sitter Universe at $\alpha =0$, which
is a plus for our model because it appears that existing models are a
specific case of our parametrization.

It is clear from Figs. \ref{Fr0}-\ref{Fr2} that the energy density of the Universe exhibits positive behavior for all values of the model parameter $%
n $, as predicted. As we can see, the condition $n\geq \frac{1}{2}$ is
imposed by positive energy density.

The equation of state (EoS) parameter specifies the relation between
pressure and energy density. The EoS parameter is used to categorize
distinct stages of the known Universe decelerated and accelerated expansion.
The matter-dominated phase is given by $\omega =0$. In the recent
accelerated stage of evolution, $-1<\omega <-\frac{1}{3}$ depicts the
quintessence era, $\omega =-1$ denotes the $\Lambda $CDM paradigm, and $%
\omega <-1$ is the phantom era. Fig. \ref{FE0} displays the evolution of the
EoS parameter for $n=0.55$. All three
constrained values of the model parameters are found to conform to the $%
\Lambda $CDM model for $z\rightarrow -1$ (future) and to the quintessence
Universe for $z=0$ (present). Fig. \ref{FE1} shows the curve of the EoS
parameter for $n=1.5$, and we can see the same behavior as mentioned in $%
n=0.55 $. As seen in Fig. \ref{FE2}, for $n=2.0$, it approaches the matter era
at $z=0 $ and early times.

The speed of sound parameter $v_{s}^{2}$ is a crucial measure for
understanding the stability of a DE model under perturbations. If $%
v_{s}^{2}>0$, the model is stable; otherwise, it is unstable. The expression
for $v_{s}^{2}$ is written as $v_{s}^{2}=\frac{\partial _{z}p}{\partial
_{z}\rho }$ \cite{speed}. The evolution of $v_{s}^{2}$ for our Hubble
parametrization for $n=0.55$ is shown in Fig. \ref{Fv0}. It is clear that $v_{s}^{2}<0$ for all three constrained values of
the model parameters and so the model is unstable. For $n=1.5$ and and $n=2.0$, the models are
stable ($v_{s}^{2}>0$) as shown in Figs. \ref{Fv1} and \ref{Fv2}.

The Om diagnostic is a simple tool that just requires the first-order time
derivative of the scale factor, which involves the Hubble parameter. It is
used to differentiate the $\Lambda $CDM model in order to clarify
alternative DE theories. The $Om(z)$ diagnostic is derived for a spatially
flat Universe as \cite{Sahni},

\begin{equation}
Om\left( z\right) =\frac{\frac{H^{2}\left( z\right) }{H_{0}^{2}}-1}{\left(
1+z\right) ^{3}-1}.
\end{equation}

Different DE models may be characterized using $Om(z)$ behavior. The
phantom type, $\omega <-1$, is according to the positive slope of $Om(z)$,
while the quintessence type, $\omega >-1$, corresponds to the negative slope
of $Om(z)$. The $\Lambda $CDM model is represented by constant behavior of $%
Om(z)$. In Fig. \ref{FOm}, the $Om(z)$ shows a negative slope, indicating
accelerated expansion and quintessence-like behavior.

\section{Conclusion}

\label{sec7}

With this study, we investigated the accelerated expansion phase of the
Universe in $Q^{n}$ gravity, where $Q$ is the non-metricity scalar. In this
background, we used a model-independent approach \cite{Pacif1, Pacif2} to
parameterize the Hubble parameter $H\left( z\right) $ in Eq. (\ref{Hz}). The
newly proposed parametrization is similar to the most widely accepted $%
\Lambda $CDM model in cosmology where $\beta =0$, corresponds to a Universe
full of matter, and for $\alpha =0$, it corresponds to a Universe filled
with dark energy. Thus, the free parameter $\beta $ is an excellent
indicator of deviation of the current dark energy parametrization from the $%
\Lambda $CDM model. We also used the $\chi ^{2}$ minimization approach to
constrain the model parameters using the most recent 57 points of Hubble
data sources, 1048 points of SNe data sources, and 6 points of BAO data
sources. Figs. \ref{C1}-\ref{C3} show the constraints on the model
parameters at $1-\sigma $ and $2-\sigma $ confidence intervals using the
Hubble data sources, SNe data sources, and BAO data sources, respectively.
The best fit values of the model parameters $H_{0}$ and $\alpha $ are shown
in the Tab. \ref{tab}. Here, the likelihoods contours are perfectly suited
to Gaussian distributions. The model is also compared to the $\Lambda $CDM
model, as seen in the error bar graphs.

First, Fig. \ref{Fq}, shows the evolution of the deceleration parameter for
our Hubble parametrization. For all constrained values of the model
parameters, our parametrization displays a successful transition from the
deceleration ($q>0$) stage to the acceleration ($q<0$) stage of the Universe in recent times. The present value of the deceleration parameter is
$q_{0}=-0.406^{+0.079}_{-0.072}$, $q_{0}=-0.461^{+0.174}_{-0.163}$, and $q_{0}=-0.320^{+0.096}_{-0.088}$ according to the Hubble,
SNe and BAO data sources, respectively. These results are found to be in excellent agreement with previous findings reported by various researchers using diverse approaches \cite{Koussour2, Hernandez}. Nair et al. \cite{Nair} conducted a comprehensive analysis that combines multiple cosmological probes, including standard candles, rulers, and cosmic clocks. The authors derived values for the deceleration parameter that closely align with our results. Specifically, they reported $q_{0}$ values of $q_{0}=-0.371 \pm 0.023$, $q_{0}=-0.526 \pm 0.028$, and $q_{0}=-0.796$ using three different parametrizations of the deceleration parameter. Wang et al. \cite{Wang} investigated the implications of a logarithmic scalar field dark energy model and its influence on the evolution of black hole masses in the cosmos. The authors performed a rigorous analysis of observational data, combining cosmic chronometers, BAO, and SNe to constrain the model parameters. Their analysis yielded a deceleration parameter value of $q_{0}=-0.43^{+0.06}_{-0.06}$. Then, to shed some light on the
influence of the functional form of $f\left( Q\right) $, we have examined
three cosmological models based on the value of the parameter $n$,
especially $n=0.55$, $n=1.5$ and $n=2.0$. For all three models, we found the energy density to be positive, as expected. In addition, we
investigated the behavior of the EoS parameter of the three models and it
behaves as the quintessence model of DE. Gadbail et al. \cite{Par5} focused on the parametrization of the deceleration parameter within the framework of $f(Q)=\alpha Q^n$ gravity. The authors selected $n = 1.2$ and $\alpha = -0.01$, observing behavior similar to that of quintessence in our three models. These findings are consistent with our results. Using a similar approach, in Ref. \cite{Par5}, the authors investigated a model of $f(Q)= Q+m Q^n$ with three values of the parameter $n$. For $n < 2$, they observed an accelerating phase with $\omega < -1/3$, while for $n \geq 2$, they found a decelerating phase with $\omega > -1/3$. Recently, Koussour et al. \cite{sp} investigated the constant sound speed concept and its thermodynamic implications in the framework of the $f(Q)=\alpha Q^n$ power-law model. Their approach differed from ours, as they utilized observational data ($Hubble+SNe+BAO$) to constrain the model's parameters, yielding the values $n=0.5034^{+0.0014}_{-0.0020}$ and $\alpha =-0.797^{+0.010}_{-0.010}$. Finally, the $Om(z)$
diagnostic for our model shows a negative slope, indicating also accelerated
expansion and quintessence-like behavior. As a result, we conclude that the
current parameterized model agrees well with recent observations. Lastly, in future studies, it would be a good idea to apply this parametrization to another gravity scenario.

\textbf{Data availability} There are no new data associated with this
article.


\end{document}